\documentclass[a4paper,preprint,showpacs,amssymb,pre,superscriptaddress]
              {revtex4}
\usepackage{graphicx}

\begin{document}

\def\prl{{\em Phys. Rev. Lett. }}
\def\prb{{\em Phys. Rev. B }}
\def\jap{{\em J. Appl. Phys. }}
\def\ajp{{\em Am. J. Phys. }}
\def\nimb{{\em Nucl. Instr. and Meth. Phys. B }}
\def\apl{{\em Appl. Phys. Lett.}}

\title{Brownian motors: Joint effect of non-Gaussian noise and
time asymmetric forcing}

\author{Raishma Krishnan}\email{raishmakrishnan2000@yahoo.co.in}
\author{Horacio S. Wio}\email{wio@ifca.unican.es}
\affiliation{Instituto de F\'{\i}sica de Cantabria, Universidad de
Cantabria-CSIC\\ E - 39005 Santander, Spain}

\begin{abstract}
Previous works have shown that time asymmetric forcing on one hand,
as well as non-Gaussian noises on the other, can separately enhance
the efficiency and current of a Brownian motor. Here, we study the
result of subjecting a Brownian motor to both effects
simultaneously. Our results have been compared with those obtained
for the Gaussian white noise regime in the adiabatic limit. We find
that, although the inclusion of the time asymmetry parameter
increases the efficiency value up to a certain extent, for the
present case this increase is much less appreciable than in the
white noise case. We also present a comparative study of the
transport coherence in the context of colored noise. Though the
efficiency in some cases becomes higher for the non-Gaussian case,
the P\'eclet  number is always higher in the Gaussian colored noise
case  than in the white noise as well as non-Gaussian colored noise
cases.
\end{abstract}
\pacs{02.50.Ey, 05.40.-a, 05.45.-a}

\maketitle

\section{Introduction}

In recent years \textit{noise induced transport} by Brownian motors
or ``ratchets" \cite{review1} have attracted the attention of an
increasing number of researchers. Such interest was motivated by
their biological interest as well as for its potential technological
applications. The pioneering works which included a built-in
ratchet-like bias together with correlated fluctuations, were
followed by works including and studying different aspects such as
tilting and pulsating potentials, velocity inversions, etc. Thorough
reviews in this area \cite{reiman} indicate the biological and/or
technological motivation for the study of ratchets as well as the
state of the art. A large part of the contemporary literature points
towards different means to improve the efficiency or the current in
a ratchet system.

There are current studies on the role of non-Gaussian noises on some
noise-induced phenomena \cite{qRE1,qRE1p,qRE2,qruido1,qruido2,wio3}
showing the possibility of strong effects on the system's response.
Such a noise source is based on the nonextensive statistics
\cite{tsallis} with a probability distribution that depends on $q$,
a parameter indicating the departure from Gaussian behavior: for $q
= 1$ we have a Gaussian distribution, and different non-Gaussian
distributions for $q > 1$ or $q < 1$. In \cite{qruido4} it was shown
that the effect of such a non-Gaussian colored noise can strongly
enhance the transport properties of Brownian motors.

Another line of work, also pointing towards efficiency enhancement,
can be found in \cite{rk1,rk2,rk3,ai}. The system considered in
these works include both time and space asymmetries \cite{ajdari},
thereby finding  a range of parameters where a remarkable efficiency
enhancement is obtained.

In the present work we explore the possibility of mixing up both
previous enhancement methods. That is, to consider the system used
in \cite{rk1}, but subject to a non-Gaussian colored noise source as
used in \cite{qruido4}. As we indicate latter, it is in principle
possible to try an analytical-like approach using an effective
Markovian approximation \cite{qRE1,wio3}. However, here we have
chosen to make an extensive numerical analysis of the problem.
Moreover, as was shown in \cite{wio3,qruido4}, the enhancement is
found to occur only for $q>1.0$, and for this reason we restrict our
analysis only to such a finite range of the parameter $q$.

We also study the transport coherence in this model by analyzing the
P\'eclet number $Pe$. This is a dimensionless number relevant in the
study of transport phenomena, defined as the ratio of the rate of
advection of a physical quantity by the flow to the rate of
diffusion of the same quantity driven by an appropriate gradient
\cite{landau}. In our present case $Pe$ is defined as the ratio of
current to the effective diffusion coefficient in the medium and is
expressed as $Pe = \frac{l {v}}{D_{eff}},$ where $l$ is a
characteristic length (in this case the length period=2$\pi$), $v$
the velocity, and $D_{eff}$ the (effective) diffusion coefficient.
The transport of a Brownian particle is always accompanied by the
spreading of fluctuations, namely, a diffusive spread, in the physical space
at a fixed time, and the effectiveness of transport is affected by
such a diffusive spread. The Brownian particle takes a time $t=l/v$
to traverse a distance $l$ with a velocity $v$ and the diffusive
spread of the particle during the same time is given by $<(\Delta
x)^2>=2D_{eff} t $. Hence, the criterion to have a reliable
transport is that $<(\Delta x)^2>=2D_{eff} t < l^2$. This implies
that $Pe=lv/D_{eff} >2$ for coherent transport \cite{pe}. The value
of $Pe$ depends on the characteristic length scale $l$ of the
system.

Our results indicate that, even though we observe an enhancement in
efficiency, in general it is smaller than for the white noise case.
However, there are regions of parameters (like noise intensity, noise
correlation time, departure from Gaussian behavior) that shows a
remarkable optimization of the transport properties, that could have
a strong interest for technological applications. Our results on
$Pe$ shows that the inclusion of colored noise causes an enhancement in
$Pe$ as compared to that in the white noise case, though, the value
of $Pe$ is not high enough to make the net transport coherent over a
wide range of parameter space. Also, P\'eclet  number is always
higher in the Gaussian colored noise case  than in the non-Gaussian
case at variance to the behavior of efficiency.

In the next section we describe the model, the noise source, and the
simulation procedure. In section III  we present our results and discuss
them in detail. The last section includes some
general conclusions and indications for future work.

\section{Model}

We consider the motion of an overdamped Brownian particle subjected
to a temporally asymmetric adiabatic forcing in the presence of a
random noise source. The Langevin equation of motion of such a
particle is
\begin{equation}
\frac{dx}{dt}= f(x) + F(t) - L + \zeta(t). \label{leq}
\end{equation}
The periodic ``ratchet" potential  that we have adopted is $V(x)= -V_0 \sin(x)
- \mu/4 \sin(2x)$, with the force field given by $f(x)=-dV(x)/dx$. Here,
$\mu$ is a measure of the asymmetry in the periodic potential
and we have scaled $V_0$ to be unity. The parameter $L$ in Eq.~(\ref{leq})
is the external load force and $\zeta(t)$, that in the original formulation
\cite{rk1,rk2,rk3} was assumed white, is a non-Gaussian colored noise
governed by the equation
\begin{equation}
\frac{d\zeta(t)}{dt}=-\frac{1}{\sigma} \frac {d
V_q(\zeta)}{d\zeta}+{\frac{1}{\sigma}\xi(t)}, \label{lan}
\end{equation}
where $\xi(t)$ is a zero mean and delta correlated ($<\xi(t)
\xi(t^\prime)> = 2 D \delta(t-t^\prime)$) Gaussian white noise. The
potential $V_q(\zeta)$ is
\begin{equation}
V_q(\zeta)=\frac{1}{\beta(q-1)} \ln \left( 1 +
\beta(q-1)\frac{\zeta^2}{2} \right),
\end{equation}
with $\beta=\sigma/D$. For $q=1$ we recover the Gaussian
Ornstein-Uhlenbeck noise, a colored noise with time correlation $\sigma$.
When $q=1$ and $\sigma=0$ we recover the Gaussian white noise case. Figure
1 in Refs. \cite{wio3,qruido4} depict the typical form of the probability
distribution function for this process.

The periodic zero-mean forcing $F(t)$ in the above
equation Langevin equation Eq.~(\ref{leq}) is  taken to be asymmetric 
in time and is given by
\begin{eqnarray}
F(t)&=& \frac{1+\epsilon}{1-\epsilon}\, F_0,\,\,\,\,\,\,\,\,\,\,\,\,
n\tau \leq \, t \, < n\tau + \frac{1}{2} \tau (1-\epsilon), \\
\nonumber &=& -F_0,\,\,\,\,\,\,\,\,\,\,\,\,\,\,\,\,\,\,\,\,\,\,
n\tau + \frac{1}{2} \tau (1-\epsilon) < \, t \, \leq (n+1)\tau .
\end{eqnarray}

\begin{figure}[hbp!]
\includegraphics[width=4.5cm]{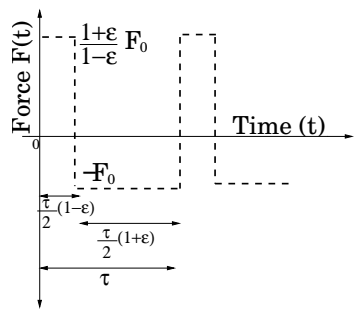} \caption{Sketch of the time
asymmetric forcing.} \label{fig:fig1}
\end{figure}

The parameter $\epsilon$ here characterizes the temporal asymmetry of the
periodic forcing, $\tau$ is the time period of the driving force
$F(t)$ and $n=0,1,..$  is an integer. We assume that $F(t)$ changes slow 
enough,
i.e., its frequency is smaller than any other
frequency related to the relaxation rate
in the problem such that the system is in a steady
state at each instant of time.  Thus we consider the adiabatic limit of forcing
and the profile of this forcing $F(t)$ is shown in Fig. 1.

Following the Stratonovich interpretation, the corresponding
Fokker-Planck equation associated to Eqs. (\ref{leq},\ref{lan}) is
\begin{eqnarray}
\frac {\partial P_q(x,\zeta,\sigma,t)} {\partial t}&=& -\frac
{\partial}{\partial x} \Bigl( [f(x)+F(t)-L+\zeta]
P_q(x,\zeta,\sigma,t) \Bigr) \nonumber \\ & + &
\frac{\partial}{\partial \zeta}\left( \frac{1}{\sigma}
\left(\frac{d}{d\zeta}V_q(\zeta)\right) P_q(x,\zeta,\sigma,t)\right) +
\frac{D}{\sigma^2} \frac{\partial^2}{\partial \zeta^2}
P_q(x,\zeta,\sigma,t).
\end{eqnarray}

Since we are interested in the adiabatic limit it is possible 
to obtain an expression
for the probability current density $j$ in the presence of a
constant external force $F_0$ in the limit $q=1$ and $\sigma=0$. 
The expression
is given by
\begin{eqnarray}
j&=&\frac{1 - \exp {[\frac{-2\pi F_0 }{k_BT}]}}
 {{\int_{0}^{2 \pi}dy I_-(y)}},
 \end{eqnarray}
 where  $I_-(y)$ is given by
 \begin{equation}
I_-(y)=\exp\,\left[\frac{-V(y)+ F_0y}{k_BT}\right]
\int_{y}^{y+2\pi}dx \exp\,\left[\frac{V(x)-F_0x}{k_BT}\right].
\end{equation}
and $k_BT$ corresponds to the thermal noise which corresponds 
to $D$ in the present case. It may be noted that for $\mu=0$, $j(F_0)
\neq -j(-F_0)$ for $\phi \neq 0\,,\,\pi$. This asymmetry
ensures rectification of current for the rocked ratchet
even in the presence of spatially symmetric potential.

The net current in the system arises due to the effect of the
non-Gaussian noise as well as the zero mean time asymmetric forcing.
The expression for the time averaged current due to the time
asymmetric forcing alone can be separated into two parts
\begin{equation}
 \langle j \rangle_f=j^+ + j^-,
\label{totj}
\end{equation}
where
\begin{eqnarray}
j^+ & = & \frac{1}{2}(1-\epsilon)\,
j(\frac{1+\epsilon}{1-\epsilon}F_0) , \nonumber \\
j^- & = & \frac{1}{2}(1+\epsilon)\,j(-F_0).
\end{eqnarray}
Here, $j^+$ is the fraction of current that flows during the
interval of time $(1-\epsilon) \frac{\tau}{2}$ when the external
driving force is $\frac{1+\epsilon}{1-\epsilon} F_0$, and $j^-$ is
the fraction of current that flows during the time period when the
external driving force is $-F_0$. The difference between the total
current, $J$, and the net current $ \langle j \rangle_f $ from the 
time asymmetric forcing term, i.e.,  $j_c=J- \langle j \rangle _f$ 
comes from the colored non-Gaussian noise which in
turn depends on $q$ and $\sigma$.

The contribution to the input energy also comes from both the
colored noise and the time asymmetric forcing i.e.,  $E_{in} = E_{in-c} +
E_{in-f}$. The input energy per unit time due to the time asymmetric
forcing, $E_{in-f}$, can be expressed by \cite{kamegawa}
\begin{eqnarray}
E_{in-f}=\frac{1}{2} F_0 \left[ \left( \frac{1+\epsilon}{1-\epsilon}
\right) j^+ - j^- \right].
\end{eqnarray}
The contribution to the input energy from the non-Gaussian colored
noise is given by \cite{qruido4}
\begin{eqnarray}\label{enav}
E_{in-c}=\lim_{t \rightarrow 0} \frac{1}{t}\int_0^{t} \zeta(t)
\frac{dx}{dt} dt = \langle -V^\prime(x)\zeta \rangle + \langle
\zeta^2 \rangle ,
\end{eqnarray}
where $\langle .. \rangle$ indicates ensemble averaging. Numerical
simulations show that the time average of $V'(x(t)) \zeta(t)$ (i.e., 
$\langle V'(x(t)) \zeta(t) \rangle$ ) is
always several orders of magnitude smaller than $\langle \zeta(t)^2
\rangle$ and hence such a term could be neglected. The quantity
$\langle \zeta(t)^2 \rangle$ can be approximated by
$2D/{\sigma(5-3q)}$ \cite{qruido4}. Thus the expression for the
contribution to the input energy from the non-Gaussian colored noise
term is approximately given by
\begin{eqnarray}
E_{in-c} \cong \frac{2D}{\sigma(5-3q)}.
\end{eqnarray}
It is worth to remark that this expression is valid only for finite
values of $\sigma$ and hence is not applicable in the presence of an
uncorrelated non-Gaussian noise source ($\sigma=0 $ and $q \neq 0$).

The most common way of defining efficiency is by attaching a load
force $L$ in a direction opposite to the direction of current in the
ratchet \cite{eff-load}. The overall potential experienced by the
Brownian particle is then given by $V(x)= - V_0 \sin(x) - (\mu/4)\,
\sin(2x) + x L$. In order to reduce the number of parameters, in
what follows we adopt the asymmetry parameter, $\mu = 0$ or, 
in other words, the potential
that the particle experience is a simple sinusoidal potential.

Within the  operating range of the load, $0<L<L_s$, the particles move
in a direction opposite to the direction of the applied load force
thereby storing energy. $L_s$ indicates a threshold value (called
stopping force) such that for $L = L_s$ the current is zero. The
average work done over a period (i.e. power) is given by
\begin{eqnarray}
E_{out}=\frac{1}{2} L \, J,
\end{eqnarray}
where $J$ is the total current \cite{rk1,rk2}.

The thermodynamic efficiency for energy transduction is given by
\begin{eqnarray}
\eta=\frac{E_{out}}{E_{in}}\,.
\end{eqnarray}
As has been explained in \cite{rk1}, at low temperature, when the
thermal energy is smaller than the modulation amplitude of the
potential, $k_B T < V_0$, a significant current value can only arise
when $F_0$, the external bias, is larger than a critical value which,
in the present case, is $1$ \cite{riskin}. When $F_0 < 1$, barriers
exists in both directions and hence there is no current. A
significant current can flow in the forward direction only when
$\frac{1+\epsilon}{1-\epsilon} F_0 >1$. Thus when
$\frac{1-\epsilon}{1+\epsilon} < F_0 < 1$, a unidirectional current
exists in the ratchet due to the temporally asymmetric periodic
forcing. In the present work the value of the zero mean external
bias $F_0$ lies in the range $0.11 < F_0 < 1$.

\section{Numerical Details}

We have numerically simulated Eqs. (\ref{leq},\ref{lan}) using the
Heun algorithm in the adiabatic limit for the external zero mean
time asymmetric forcing (that is, the system 
follows the external forcing). The ratchet system 
is evolved for $10^2$ cycles of forcing,
$t=10^2 {\tau}$, where ${\tau}=10^4$ is the time period of forcing
and the adopted time step is $10^{-2}$. This time period of external
forcing is chosen to be large enough to ensure that we are in the
limit of adiabatic forcing. We have done ensemble averages over
$10^3$ trajectories  and have calculated the current, diffusion, the
efficiency and P\'eclet number, through numerical simulations. We
have computed $v$, the net velocity, and $D_{eff}$, the effective
diffusion coefficient describing fluctuations around the average
position of the particles due to both the colored non-Gaussian noise
as well as time asymmetric forcing and the P\'eclet number 
using the expressions
\begin{equation}
v = \Big{\langle} \, \frac{x(t)-x(t_0)}{t-t_o} \, \Big{\rangle} = 2 \pi J,
\label{gencur}
\end{equation}
and
\begin{equation}
D_{eff} = \lim_{t \to \infty}
\frac{1}{2t} \left [ \langle {x^2(t)}  \rangle - {\langle x(t)
\rangle} ^2 \right]
\end{equation}
where $\langle \dots \rangle$, as in Eq. (\ref{enav}), denotes
ensemble averaging.
The  P\'eclet number is given by
\begin{equation}
Pe = \frac{l {v}}{D_{eff}},
\end{equation}
here $l$ is a characteristic length (=2$\pi$), $v$
the velocity, and $D_{eff}$ the (effective) diffusion coefficient.
All the physical quantities are taken in dimensionless units.

\section{Results and Discussions}
\begin{figure}[hbp!]
\includegraphics[width=7cm,angle=-90]{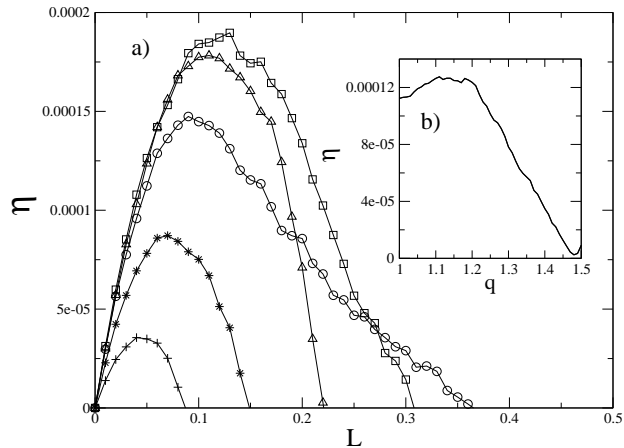}
\caption{(a) Efficiency as a function of load for $F_0=0.1$,
$\epsilon=0.8$, $D=0.1$ and $\sigma=0.1$ for different $q$ values:
$q=1.0$ ($\circ$), $q=1.1$ ($\square$), $q=1.2$ ($\diamond$), $q=1.4$
($+$), $q=1.5$ ($\ast$). The inset (b) shows the plot of
efficiency with $q$ along the load value $L=0.05$.} \label{fig:fig2}
\end{figure}

We start by plotting in Fig.~\ref{fig:fig2}(a), and for different
$q$ values, the efficiency as a function of load in presence of a
temporally asymmetric forcing with $F_0=0.1$ and $\epsilon=0.8$. The
adopted noise strength value was $D=0.1$ with a correlation time
$\sigma=0.1$. We see a non-monotonous dependence of efficiency with
load. For a fixed $q$ value, the efficiency increases with $L$,
reaches a maximum value and then decreases. We observe that as we
depart from Gaussian behavior, i.e., when $q$ departs from $1$
($q>1$), the efficiency is found to increase up to a certain value
of $q$ ($q=1.2$), and then it decreases when further increasing $q$.
To be more specific, we find $\eta_{q=1.1} > \eta_{q=1.2} >
\eta_{q=1.0}$. Also, the presence of non-Gaussian terms ($q >1$)
reduces the range of operation of the ratchet. In the inset of the
figure we show the behavior of efficiency with $q$ for the load
value $L=0.05$. The load was chosen such that there was a finite
efficiency value for all $q$. From the inset (b) we see that there
is a small increase in efficiency up to $q=1.2$ followed by a strong
decrease. Thus there is a non-monotonous behavior of efficiency with
$q$ as we depart from the Gaussian case.

\begin{figure}[htp!]
\includegraphics[width=7cm,angle=-90]{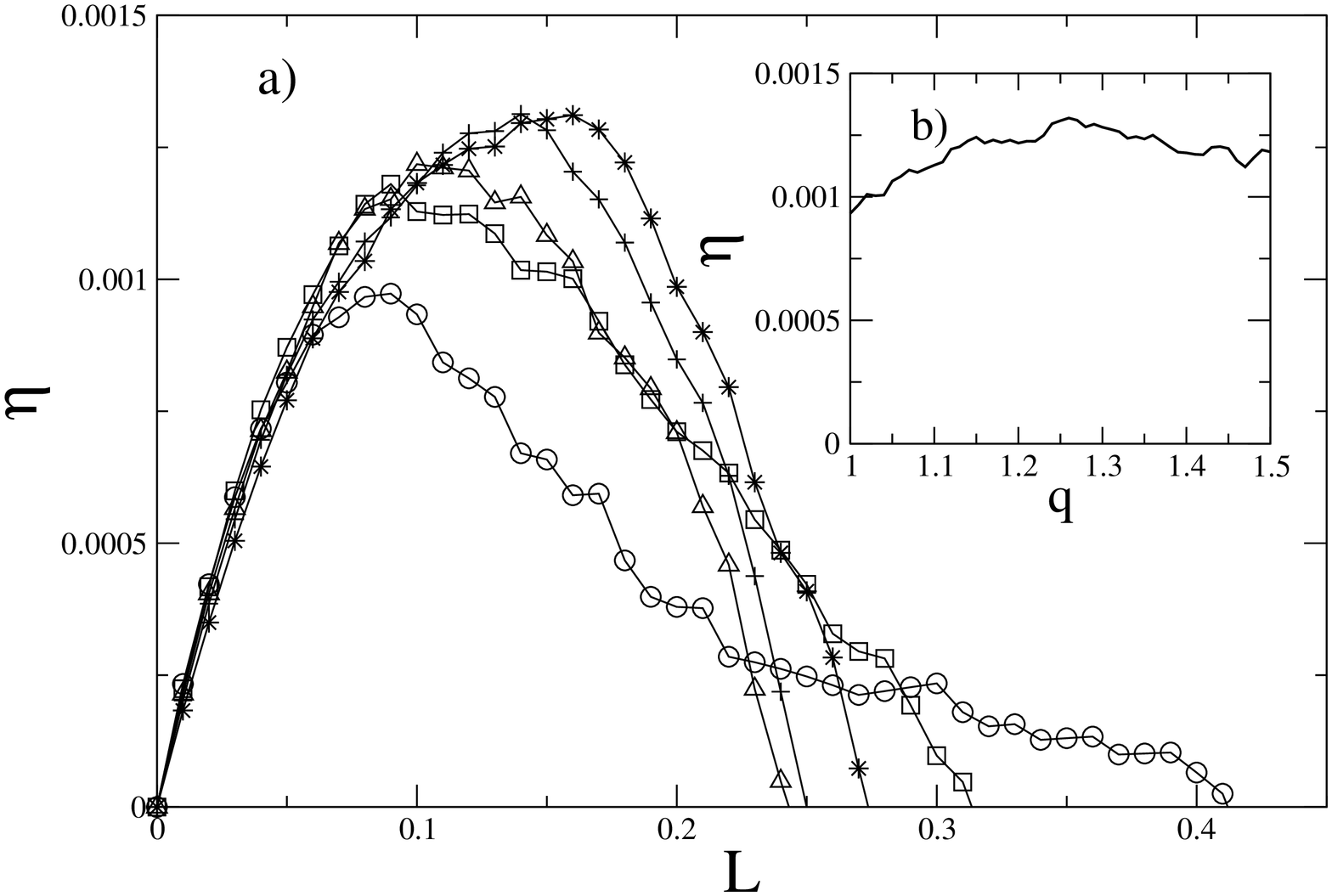}
\caption{(a) Efficiency as a function of load for $F_0=0.1$,
$\epsilon=0.8$, $D=0.1$ and $\sigma=1.0$ for different $q$ values:
$q=1.0$ ($\circ$), $q=1.1$ ($\square$), $q=1.2$
($\vartriangle$), $q=1.4$ ($+$), $q=1.5$ ($\ast$). The inset (b) shows
the plot of efficiency with $q$ along the load value $L=0.1$.}
\label{fig:fig3}
\end{figure}

In Fig.~\ref{fig:fig3} (a) we plot the efficiency as a function of
load for the same noise strength value, $D=0.1$, but with a larger
correlation time $\sigma=1.0$, and all other parameters remaining
the same. We observe that there is an increase in efficiency by an
order of magnitude  with increasing correlation time $\sigma$ when
plotted as a function of load, $L$, i.e., $\eta_{max} = 1.4 \times
10^{-3} $. The inset shows the behavior of efficiency with $q$ when
the load is $L=0.1$. We observe that there is an increase in
efficiency when we depart from $q=1.0$, however, after $q \sim
1.25$, it slightly decreases, but much less markedly than in the
inset of Fig. \ref{fig:fig2} (with $D=0.1$ and $\sigma=0.1$). Also,
contrary to the behavior seen in Fig. \ref{fig:fig2},  with lower
correlation time and noise strength $D$, the efficiency in this case
increases with increasing $q$. It should be noted that the maximum
range of load is obtained for the Gaussian case. The higher $\sigma$
value helps in keeping the efficiency around a finite value. For an
intermediate value of $\sigma$ ($\sigma=0.5$),  we have seen that the efficiency value increases up to $q=1.2$, and at $q=1.3$ 
the efficiency drops to a value below the one we have at $q=1.0$, 
when plotted as a function of load (results not shown here). Beyond
$q=1.3$ the efficiency again shoots up to a value above the Gaussian
case.

\begin{figure}[hbp!]
\includegraphics[width=8cm,angle=-90]{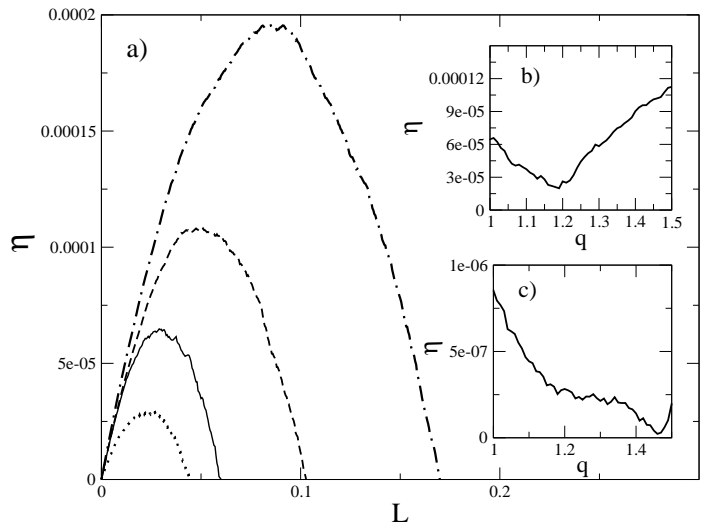}
\caption{(a) Efficiency as a function of load for $F_0=0.1$,
$\epsilon=0.8$, $D=1.0$ and $\sigma=1.0$ for different $q$ values:
$q=1.0$ (solid line), $q=1.2$ (dotted line), $q=1.4$ (dashed
line), $q=1.5$ (dash-dotted line). The inset (b) shows the
behavior of efficiency with $q$ along the load value $L=0.03$ when
$D=1.0$ and $\sigma=1.0$. The inset (c) shows the plot of
efficiency with $q$ along the load value $L=0.003$ when $D=1.0$ and
$\sigma=0.1$. } \label{fig:fig4}
\end{figure}

Figure \ref{fig:fig4} (a) shows efficiency plotted as a function of
load for a larger noise strength value,  $D=1.0$ and larger
correlation time, $\sigma=1$ with all other parameters being the
same. We see that as $q$ is increased, the efficiency value first
decreases as $q$ goes from $q=1.0$ to $q=1.2$ and then the
efficiency starts to increase. The important point to be noted here
is that the efficiency value at $q=1.4$ is larger than that at
$q=1.0$. This indicates that a correlation time $\sigma > 0$ favors
energy transduction in presence of a non-Gaussian noise ($q > 1.0$).
The interplay of $\sigma$ and $q$ gives rise to these interesting
effects even when the noise strength $D$ is very high. The inset (b)
in this figure shows the behavior of efficiency versus $q$ along the
load value $L=0.03$. We can see that the efficiency decreases with
increasing $q$ though there is a slight increase after $q=1.2$.
However, this efficiency increase is much smaller than for the
Gaussian colored noise case, $q=1$. Again, when $D=0.5$ and
$\sigma=1.0$ the maximum efficiency is obtained when $q=1.0$. The
inset (c) in this figure shows the plot of $\eta$ vs $q$ for $D=1.0$
and $\sigma=0.1$ for the load value $L=0.003$.  We can see that the
efficiency decreases with increasing $q$, reaches a minimum and then
there is a small increase for $q=1.5$. The values are so small that
the net efficiency value is almost negligible and hence we present
here as an inset only the $\eta$ vs $q$ plot for which the appropriate 
load value has been chosen from the $\eta$ vs $L$ plot.

However we find that the maximum value for efficiency and the range
of load is obtained for the Gaussian case, i.e. when $q=1$. As $q$
increases departing from the Gaussian case, the efficiency decreases
up to $q=1.4$ and then increases for $q=1.5$, though this increase
is much smaller than for $q=1.0$. Thus we can conclude that an
increase in noise strength makes the system completely random and
there is no directed transport possible in this range of parameters.

\begin{figure}[htp!]
\includegraphics[width=8cm,angle=-90]{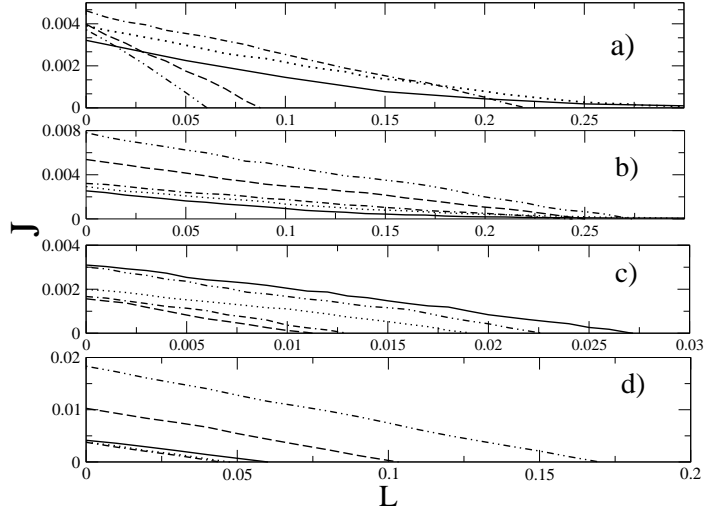}
\caption{Current as a function of load for four particular cases,
(a) $D=0.1$, $\sigma=0.1$, (b) $D=0.1$, $\sigma=1.0$, (c) $D=1.0$,
$\sigma=0.1$ and (d) $D=1.0$, $\sigma=1.0$ with  $F_0=0.1$ and
$\epsilon=0.8$. The different lines correspond to: $q=1.0$
(solid), $q=1.1$ (dotted), $q=1.2$ (dash-dotted), $q=1.4$
(dashed), $q=1.5$ (dash-double dotted)).} \label{fig:fig5}
\end{figure}

In Fig.~\ref{fig:fig5} we plot curves of current as a function of
the load for the four cases of efficiency discussed above with
different $q$ values. The figure caption gives the details of the
values chosen for $D$ and $\sigma$.  As expected, the current in the
ratchet system decreases with an increase of the load force, and
beyond a certain load value (the stopping force) the current starts
to flow in the same direction as the load. The value of current,
obtained in the presence of colored non-Gaussian noise, is much
higher than that obtained in the presence of white noise. This is
one of the main results to be stressed in our present work. From the
curves we also see that except for the case of high noise strength
and low correlation time, i.e., $D=1.0,\, \mbox{and} \sigma=0.1$,
the value of current increases with an increase in $q$. The
interplay of high noise strength and non-Gaussian behavior causes
the current to be less than in the Gaussian case value. For the case
$D=1.0$ and $\sigma=1.0$ the value obtained for the current was very
high. In fact, the value of current decreases with $q$ up to $q =
1.3$ and increases afterwards. However, we will not analyze this
parameter range as a noise source with a large $\sigma$ value
reduces the extent of randomness in the system.

\begin{figure}[hbp!]
\includegraphics[width=8cm,angle=-90]{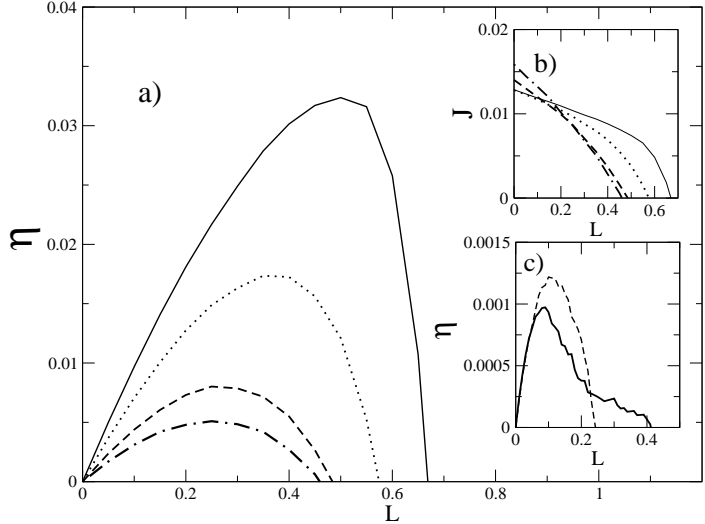}
\caption{(a) Efficiency as a function of load for $D=0.1$,
$\sigma=1.0$ and with $F_0=0.1$ and $\epsilon=0.9$. The inset (b)
shows the behavior of current for the same set of parameters. The
different lines correspond to: $q=1.0$ (solid), $q=1.2$ (dotted),
$q=1.4$ (dashed), $q=1.5$ (dash-dotted)). The inset (c) shows the
behavior of $\eta$ with load for $\epsilon=0.8$ and for $q=1.0$
(solid lines) and $q=1.2$ (dashed lines).} \label{fig:fig6}
\end{figure}

In Fig.~\ref{fig:fig6} (a) we plot the variation of efficiency with
load for the optimum value of system parameters, namely
$\sigma=1.0$, $\epsilon=0.9$, $F_0=0.1$ and $D=0.1$. We see an
enhancement in the efficiency for this specific range of parameters.
As noted before, the efficiency keeps increasing when increasing
$\sigma$. However, as indicated above, large values of $\sigma$
implies a more deterministic behavior of the system \cite{reentr}
forcing us to work with low $\sigma$ values. The inset (b) shows the
behavior of current for the same set of parameters. The obtained
value of current is high and it increases when we increase the
departure from Gaussian behavior. But the stopping value of the load
is high when noise source is colored and Gaussian.

\begin{figure}[hbp!]
\includegraphics[width=7cm,angle=-90]{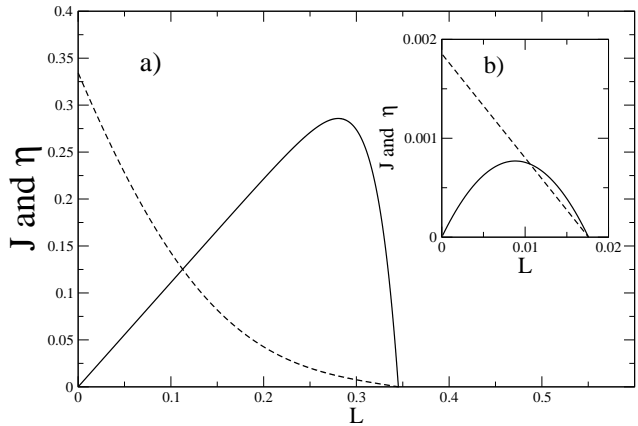}
\caption{(a) Efficiency (solid line) and current (dotted line) as a
function of load for the white noise case with $D=0.1$. The current
is multiplied by a factor of 100 to scale with $\eta$.  The inset (b)
shows $\eta$ and current as a function of load for the case when
$D=1.0$. The other parameter values chosen in the plot are
$F_0=0.1$ and $\epsilon=0.8$.} \label{fig:fig7}
\end{figure}

To grasp the relevance of the time asymmetry parameter $\epsilon$,
we checked the value of efficiency both with and without $\epsilon$.
From the obtained results we could conclude that finite $\epsilon$
values contribute to the enhancement of efficiency even for
correlated and non-Gaussian noise. However, this increase is still
very small when compared with the increase in efficiency observed in
the white noise case. We find that in order to have appreciable
efficiency values, a large $\epsilon$ ($\epsilon \geq 0.6$) is
needed. This means that the asymmetry in the temporal forcing has to
be high which is in contrast with the white noise case.

We also see that efficiency is higher when $\epsilon=0.9$ than when
$\epsilon=0.8$ by almost an order of magnitude for the same $q$
value. Thus for comparison, in the inset (c) of  Fig.~\ref{fig:fig6}
we plot $\eta$ vs $L$ when $\epsilon=0.8$ with other parameters
being the same. We represent here only two cases of efficiency value
corresponding to $q=1.0$ and $q=1.2$. A similar behavior is seen for
all other parameter ranges. We can clearly see that there is an
enhancement in efficiency with higher $\epsilon$ values with a
larger value being for the case when $q=1.0$. Thus we can conclude
from Fig.~\ref{fig:fig6} that the presence of correlation in noise
reduces the efficiency value by a large amount even when the
presence of barriers in one direction disappears due to the
temporally asymmetric forcing parameter $\epsilon$.

In order to compare with the known results for the white noise case
\cite{rk1,rk2,rk3}, in Fig. \ref{fig:fig7} we present the efficiency
and the current in presence of white noise, as a function of load,
and for two noise strength values (a) $D=0.1$ and (b) $D=1.0$. We
see that an increase in $D$ causes the value of efficiency and also
the range of load to be  drastically reduced. The order of magnitude
of current is almost the same and lies between $0.0015 \sim 0.002$
for the case when $D=1.0$ and is between $0.003 \sim 0.0035$ for the
case when $D=0.1$.

\begin{figure}[hbp!]
\includegraphics[width=7cm,angle=-90]{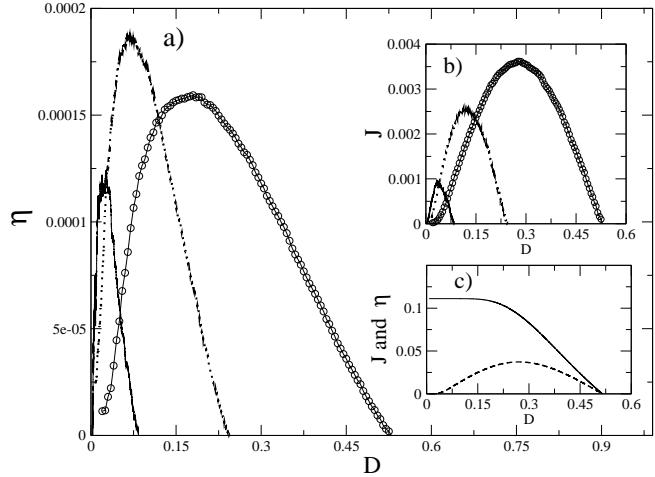}
\caption{(a) Efficiency vs D for $\epsilon=0.8$, $L=0.1$ and
$\sigma=0.1$ for different $q$ values: $q=1.0$ ($\circ$), $q=1.2$
(dotted), $q=1.4$ (solid). Inset (b) shows the behavior of current
with $D$ for the same set of parameters. In the inset (c) we show
the behavior of current (dashed) and efficiency (solid) for the
white noise case as a function of noise strength $D$. Here the
current is multiplied by a factor of 10 to scale with the efficiency
value in the plot.} \label{fig:fig8}
\end{figure}

In Fig.~\ref{fig:fig8} (a) we plot the efficiency as a function of
noise strength $D$ with $\epsilon=0.8$, $L=0.1$, and $\sigma=0.1$,
for different $q$ values. The  inset (b) shows the dependence of the
current with $D$ for the same set of parameters. We can see that
both the current and efficiency show a peak as a function of $D$ for
all range of parameters implying that noise facilitates energy
transduction. To be more specific, both the efficiency and current
are zero up to a certain value of $D$, beyond which there is a
finite current and efficiency with a peak value for a particular $D$
in between and then with further increase in $D$ there is a
reduction in current and efficiency. This is at variance with the
behavior of efficiency in presence of white noise for both
sinusoidal and sawtooth potential \cite{rk1,rk2,rk3}.

The existence of a peak in efficiency with $D$ can be understood by
plotting the curves of input energy and current as a function of the
noise intensity $D$ for the corresponding set of parameters. On
increasing $D$, the current starts to rise to a larger value as
compared to zero temperature value, and there is also an increase in
efficiency. However, when $D$ increases beyond a certain value,
there will be a contribution to the current in both directions, due
to the overcoming of the potential barrier in either directions,
implying a reduction of the net current. As a result, the efficiency
will in turn has to decrease in the high temperature regime.

The shift in the efficiency maximum when departing from Gaussian regime
is apparent from the plot. Such a maximum in $\eta$ has
a non-monotonic behavior, with the largest value occurring for
$q=1.2$, while the current maxima, as a function of $D$, has a
monotonic nature. The maximum value of the current decreases when
departing from Gaussian regime as is seen in inset (b) of this figure.

To have a comparison with the white noise case, in Fig.
\ref{fig:fig8} we have included an inset (c) where we show the
behavior of efficiency and current with a sinusoidal potential,
subjected only to white noise. We see that for this case, though the
current has a maximum as a function of $D$ the efficiency does not
show a maximum. Only a fine tuning of the different parameters could
yield a maximum in the current and efficiency with $D$. This is in
contrast with that for the colored noise case where it is possible
to obtain a peak in $\eta$ and $J$ with $D$ for both Gaussian and
non-Gaussian noises, though the peak value of efficiency and current
are very small.

\begin{figure}[htp!]
\includegraphics[width=7.5cm,angle=-90]{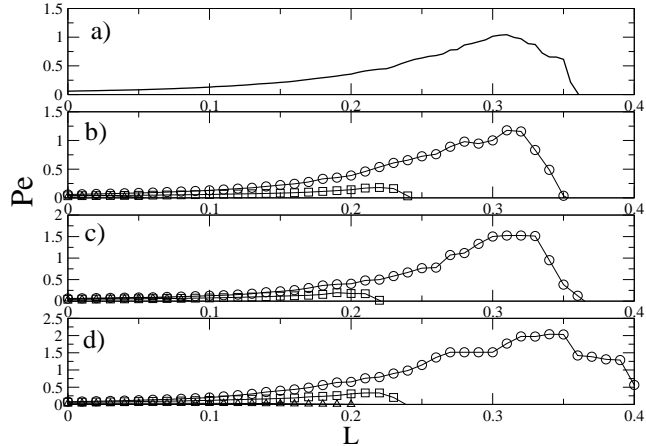}
\caption{ Plot of $Pe$ as a function of load for four
different cases of interest: (a) white noise case, (b) $D=0.1$ and
$\sigma=0.01$, (c) $D=0.1$ and $\sigma=0.1$, (d) $D=0.1$ and
$\sigma=1.0$. Other parameters are $\epsilon=0.8$ and $q=1.0$ ($\circ$),
$q=1.2$ ($\square$) and $q=1.4$ ($\vartriangle$)).}
\label{fig:fig9}
\end{figure}

Finally we analyze the transport coherence by calculating the
P\'eclet number ($Pe$) and make a comparison with the white noise
case. Fig.~\ref{fig:fig9} shows the behavior of $Pe$ as a function
of the load for the cases of interest in this work. It is apparent
that the inclusion of colored noise causes an enhancement in $Pe$ as
compared to that in the white noise case, though the value of $Pe$
is not high enough to make the net transport coherent.

From the above plot we can also observe that though the value of
$Pe$ is not high to make the transport coherent, the best value of
$Pe$ is always obtained in the Gaussian colored noise case. In other
words, though the efficiency in some cases becomes higher for the
non-Gaussian case, the P\'eclet  number is always high in the
Gaussian case. The inclusion of non-Gaussianity makes the transport
less and less coherent.

A final observation from the above plot is that the $Pe$ approaches
a value of $2$ for a small range of load values for the case when
$D=0.1$ and $\sigma=1.0$. Thus, as in the case of efficiency, an
increase in $\sigma$ makes the transport to be coherent. We can
conclude that in the Gaussian colored noise case it is possible to
have a larger transport coherence than in the white noise case
(i.e., with the inclusion of $\sigma$), due to the larger current.
However, the efficiency value reduces with the inclusion of
$\sigma$.

\section{Conclusions}

We have studied the combined effect of a non-Gaussian colored noise
source and a zero mean temporal asymmetric rocking force on crucial
quantities of the ratchet system, namely the current, efficiency and
P\'eclet number. We have seen that though time asymmetric forcing
helps to increase the efficiency, such an increase is not as
appreciable as in the white noise regimes. However, we have found
that there exists a parameter regime for noise intensity, noise
correlation time, and departure from Gaussian behavior, showing a
notable optimization of the transport properties, that could have a
strong interest for technological applications.

Finally, a direct analytic approach, exploiting an effective Markovian
approximation \cite{qRE1,qRE1p} is possible, but highly cumbersome
from a numerical point of view. Another possibility that we are 
trying to implement is a perturbation approach for small $\sigma$ 
and for $q \sim 1$ (i,e, $|q - 1| \ll 1$). We expect that such an 
approach will reveal some details about 
the behavior of the system in relation to the role of the 
different parameters on efficiency
and current. In particular, we expect it could clarify how the
two enhancement methods interacts leading to a poor enhancement 
when combined. Such studies will be the subject of a
forthcoming work \cite{futur}.

\begin{acknowledgments}
RK acknowledges the award of a Postdoc Fellowship from MEC, Spain.
HSW acknowledges financial support from MEC, Spain, through
CGL2007-64387/CLI.
\end{acknowledgments}


\begin{thebibliography}{0}

\bibitem{review1} R. D. Astumian, Science {\bf 276}, 917 (1997).

\bibitem{reiman} P. Reimann, Phys. Rep. {\bf 361}, 57 (2002), P.
H\"{anggi} and F. Marchesoni, Rev. Mod. Phys. {\bf 81}, 387 (2009);
see also A. M. Jayannavar, cond-mat 0107079.

\bibitem{qRE1} M.A. Fuentes, R. Toral and H.S. Wio, Physica A {\bf
295}, 114-122 (2001); M.A. Fuentes, H.S. Wio and R. Toral, Physica A
{\bf 303}, 91--104 (2002).

\bibitem{qRE1p} M.A. Fuentes, C. Tessone, H.S. Wio and R. Toral,
Fluctuations and Noise Letters {\bf 3}, L365 (2003).

\bibitem{qRE2} F.J. Castro, M.N. Kuperman, M.A. Fuentes and H.S.
Wio, Phys. Rev. E {\bf 64}, 051105 (2001).

\bibitem{qruido1}  H.S. Wio, J.A. Revelli and A.D. S\'anchez,
Physica D {\bf 168-169}, 165-170 (2002).

\bibitem{qruido2} H.S. Wio and R. Toral, in {\it Anomalous
Distributions, Nonlinear Dynamics and Nonextensivity}, H. Swineey
and C. Tsallis (Eds.), Physica D {\bf 193}, 161 (2004).

\bibitem{wio3} H.S. Wio, {\it On the Role of Non-Gaussian
Noises}, in {\it Nonextensive Entropy-Interdisciplinary
Applications}, M.Gell-Mann and C. Tsallis, Eds. (Oxford U.P.,
Oxford, 2003).

\bibitem{tsallis} C. Tsallis, Stat. Phys. {\bf 52}, 479 (1988);
E.M.F. Curado and C. Tsallis, J. Phys. A {\bf 24}, L69 (1991);
ibid, {\bf 24}, 3187 (1991); ibid, {\bf 25}, 1019 (1992).

\bibitem{qruido4} S. Bouzat and H.S. Wio, Eur. Phys. J. B.
\textbf{41}, 97-105 (2004); S. Bouzat and H.S. Wio, Physica A
\textbf{351}, 69-78 (2005).

\bibitem{rk1} R. Krishnan, M.C. Mahato and A.M. Jayannavar, Phys.
Rev. E \textbf{70}, 021102 (2004).

\bibitem{rk2} R. Krishnan, S. Roy and A.M. Jayannavar, J. Stat.
Mech. P04012 (2005).

\bibitem{rk3} R. Krishnan, J. Chacko. M. Sahoo and A.M. Jayannavar,
J. Stat. Mech. P06017 (2006).

\bibitem{ai} B.-Q. Ai, X. J. Wang, G. T. Liu, D. H. Wen, H. Z. Xie,
W. Chen and L. G. Liu, Phys. Rev. E \textbf{68}, 061105 (2003).

\bibitem{ajdari} A. Ajdari, D. Mukamel, L. Peliti and J. Prost,
J. Phys. I France {\bf 4}, 1551 (1994), M.C. Mahato and A.M.
Jayannavar, Phys. Lett. A {\bf 209}, 21 (1995), D.R. Chialvo, M.M.
Millonas, Phys. Lett. A {\bf 209}, 26 (1995), M.C. Mahato, T.P.
Pareek and A.M. Jayannavar, Int. J. Mod. Phys. B {\bf 10}, 3857
(1996).
\bibitem{landau} L. D. Landau and E. M. Lifshitz, Fluid Dynamics,
Pergamon Press, Oxford, 1959.

\bibitem{pe} S. Roy, D. Dan and A. M. Jayannavar, J. Stat.
Mech. P09012 (2006); R. Krishnan, D. Dan
and A. M. Jayannavar, Mod. Phys. Lett. B {\bf 19} Nos 19 \& 20, 971 (2005);
R. Krishnan, D. Dan and A. M. Jayannavar, Physica A {\bf 354},
171 (2005);  B. Linder and L Schimansky-Geier, Phys. Rev. Lett. {\bf 89},
 230602 (2002), R. Krishnan, D. Dan and A. M. Jayannavar,
Ind. J. Phys. {\bf 78}, 747 (2004),  J. A. Freund and L. Schimansky-Geier,
Phys. Rev. {\bf E60}, 1304 (1999); K. Visscher, M. J. Schnitzer
and S. M. Block, Nature {\bf 400}, 184 (1999), T. Harms and R. Lipowsky,
Phys. Rev. Lett. {\bf{79}}, 2895 (1997), M. J. Schnitzer
and S. M. Block, Nature (London) {\bf 388}, 386 (1997).

\bibitem{kamegawa} H. Kamegawa, T. Hondou and F. Takagi, Phys. Rev.
Lett. {\bf 80}, 5251 (1998); F. Takagi and T. Hondou, Phys. Rev.
{\bf E60}, 4954 (1999); D. Dan and A. M. Jayannavar, Phys. Rev. {\bf
E66}, 41106 (2002).

\bibitem{eff-load} K. Sekimoto, J. Phys. Soc. Jpn. {\bf 66}, 6335 (1997);
J. M. R. Parrondo and B. J. De Cisneros, Appl. Phys. {\bf A75}, 179 (2002).

\bibitem{riskin} H. Risken, The Fokker-Planck Equation (Springer
Verlag, Berlin 1984).

\bibitem{reentr} F. Castro, A. S\'anchez and H.S.Wio, Phys. Rev.
Lett. \textbf{75}, 1691 (1995).

\bibitem{futur} R. Krishnan and H.S. Wio, in preparation.

\end{thebibliography}
\end{document}